\documentclass[12pt,showpacs,showkeys,preprintnumbers,nofootinbib]{revtex4-1}
\pdfoutput=1
\usepackage[pdftex,dvipsnames]{xcolor}
\usepackage{xargs}
\usepackage{graphicx}
\usepackage{hyperref}
\usepackage{epstopdf}
\usepackage{color}
\usepackage{float}
\usepackage{amsmath}
\usepackage{amsfonts}
\usepackage[utf8]{inputenc}
\usepackage{mathrsfs}
\usepackage{enumerate}

\begin{document}
\title{Role of the plurality rule in multiple choices}

\author{Angelo M. Calv\~ao$^{1,2}$}

\author{Marlon Ramos$^{1}$}

\author{Celia Anteneodo$^{1,3}$}
\thanks{E-mail: celia.fis@puc-rio.br}

\affiliation{$^1$Department of Physics, PUC-Rio, Rio de Janeiro, Brazil\\
$^2$Polytechnic Institute, UERJ,  Rio de Janeiro, Brazil\\
$^3$National Institute of Science and Technology for Complex Systems, Rio de Janeiro, Brazil 
}

\begin{abstract}
People are often challenged to select one among several alternatives. 
This situation is present not only in decisions about complex issues, e.g., 
political or academic choices, but also about trivial ones, as in daily 
purchases at a supermarket. We tackle this scenario by means of the tools of statistical mechanics. 
Following this approach, we introduce and analyze a model of opinion dynamics, 
using a Potts-like state variable to represent the multiple choices, including 
the ``undecided state'', that represents  the individuals that do not make a choice. 
We investigate the dynamics over Erd\"{o}s-R\'enyi and Barab\'asi-Albert networks, 
two paradigmatic classes  with the small-world property, and we show the impact 
of the type of network on the opinion dynamics. Depending on the number of 
available options $q$ and on the degree distribution of the network of contacts, 
different final steady states are accessible: from a wide distribution of choices 
to a state where a given option largely dominates. 
The abrupt transition between them is consistent with the sudden viral 
dominance of a given option over many similar ones. Moreover, the probability 
distributions produced by the model are validated by real data. Finally, 
we show that the model also contemplates the real situation of overchoice, 
where a large number of similar alternatives makes the choice process harder 
and indecision prevail.  
	\end{abstract}

\pacs{ 
89.75.-k, 
89.65.-s, 
02.70.Uu, 
64.60.-i,  
64.60.aq, 
}

 \maketitle

\section{Introduction}

People   frequently face  diverse situations that offer  a wide choice of  options, such  
as when looking for  a  restaurant, hotel, phone model or any basic good in the supermarket.
The number of goods increases every day. 
It is estimated that approximately 50000 new products are introduced every year in the US~\cite{ers}. 
Even within each category of items, there may be many brands and item variations without differentiated attractiveness.  
This leads to the problem of facing too many choices, termed ``overchoice'' or choice overload~\cite{Toffler1970a}.  
To make  decisions in such situations can be costly, and this stressful process 
leads to  poor decisions or no decisions at all~\cite{Baumeister2006a,Herrmann2009a,schwartz2005a}. 
Then, the advantages of multiple choices can be canceled by the disadvantages of a more complicated choice process. 
In fact,  despite representing, apparently, a positive development, many options may hinder the process of choice. 
For example,  people with many purchase options tend to have more difficulty in choosing and may end up  buying nothing~\cite{Iyengar2000a}. 
Motivated by these observations, we wonder to what extent people interactions,   
leaving aside their individual psychology,  
contribute to this scenario by  introducing, for instance, conflict and frustration. 
Then, by means of a model of opinion dynamics, we investigate the distribution of adoptions made by a population 
facing a large number of choices. 
  
In modeling people's interactions, one of the basic ingredients is imitation, or social contagion. 
In fact, imitation occurs in diverse social contexts, from the dynamics of language learning to decision making. 
Depending on the questions posed, 
diverse rules of contagion, 
from simple pairwise to group interactions, 
have been proposed and studied in recent years~\cite{Galam1986a,Galam1991a,Bikhchandani1992a,Kirman1993a,Galam1997a,Deffuant2000a,Cont2000a,Challet2005a,Curty2006a,Martino2006a,Chen2005a,Holme2006a,Galam2002a,Krapivsky2003a, Shao2009a,Biswas2012a,Crokidakis2014a,Crokidakis2012a,ramos2015a,ramos2015b}. 
However, very few works deal with many 
choices~\cite{Chen2005a,Holme2006a,axelrod1997,vazquez2007non}.  
In the vast majority of opinion models,  the opinion of an agent is represented by a binary variable, 
since many questions can be tackled through the assumption of two possible (opposite) attitudes, 
e.g.,  being either favorable or unfavorable to a given choice. 
This kind of binary variable was also inspired  in the spin-1/2 Ising model, 
leading to transpose  known results from physical to social questions. 
For our present  purpose of studying multiple-choice situations,  it is natural to consider 
a Potts-like state variable that can take  several (discrete) values.

We consider that changes from one state to another are  governed, not by simple pairwise contagion, 
like in Refs.~\cite{Holme2006a,axelrod1997,vazquez2007non} but, instead, by a ``plurality'' rule~\cite{Chen2005a}. 
This is  grounded on the  idea that an individual makes the choice that is the most  popular  among its contacts. 
In fact, when we have to choose or buy something,  especially  when there are so many similar options that 
there is not a favorite one {\em a priori} and   is not feasible to examine them all, 
it is reasonable to take into account other people's preferences~\cite{Salganik2009a}.
Naturally, the closer is the person in our network of contacts, more importance we give to its  opinion, 
since nearest neighbors in the network typically have similar  interests and tastes. 
One can use that strategy not only in   trivial or daily life issues but also in major ones such as political elections 
where many candidates compete. 
For changing or adopting a new opinion, however,  a minimum of consensus amongst the contacts is necessary. 
This is expressed in a  plurality  rule, according to which an individual is persuaded to adopt the opinion 
shared by the largest number of  its nearest neighbors.

Evolution rules based on a locally dominant opinion  have been considered before, 
for instance, the majority rule for two states introduced by Galam~\cite{Galam2002a,Krapivsky2003a}. 
It was later extended to multistate opinions~\cite{Chen2005a}, 
by considering  all-to-all interactions  where  
 the individual and all its contacts adopt the same opinion of the majority at the same time. 
A variant where, instead of the local majority, the plurality opinion is considered   
was also studied~\cite{Chen2005a}. 
But in the situations we address here, the decisions are not taken in groups  or simultaneously, rather
the choices of the individuals  are affected by their knowledge of the previous choices of theirs contacts. 
Therefore, we assume that one single individual opinion changes at a time.  
Moreover, we will include the possibility of undecided people, 
which is a realistic feature, which has been taken into account  in 3-state models, as natural extensions of 
binary cases~\cite{Biswas2012a,Crokidakis2014a,Crokidakis2012a,vazquez2003,vazquez2004,mobilia2011}.  
Furthermore, as another differential, the dynamics of our model takes place in small-world networks, that, 
even if do not facilitate analytic treatment, are more realistic than regular or mean-field settings. 
As paradigms of small-world networks, we consider two classes with distinct  degree distributions: 
Erd\"{o}s-R\'enyi (ER)~\cite{Erdos1959a,Barabasi2002a} 
and Barab\'asi-Albert (BA)~\cite{Barabasi1999a}  networks.  
 The details of the model will be defined in  Sec. \ref{sec:rules}.

The model encompasses  instances where the different alternatives have similar initial attractiveness. 
Many products sold in the internet with similar qualities and prices, e.g., music albums, shoes, etc., 
are within the model scope. 
We also make the simplification that individuals can differ only in the number of contacts.  
Other heterogeneities of the agents might be also introduced in further work.
By now, we ask a basic  question: 
how a plurality  rule molds  the decision spectrum in the simplest, homogeneous, case? 

First, we  address the classical issue in this kind of problems, about whether a consensus  
state can be achieved or not, where all (or almost all) of the individuals share the same preference. 
Then, in simulations of the model, we compute the fraction of the realizations reaching consensus. 
In non-consensus situations, we analyze the distribution of adoptions. 
We also  compute other relevant quantities such  as the fraction of undecided people and the fraction of the population 
adopting the most popular alternative.  The results will be presented in Sec.~\ref{sec:results}.

\section{Plurality modeling}
\label{sec:rules}
 
A plurality rule  governs  the opinion dynamics of $N$ agents 
 interacting through their network of contacts. 
Each agent $i$, corresponding to a node in the network, has a Potts-like opinion state variable $S_i$, 
that can take the values $s_1, \dots, s_q$ representing  $q$ electable alternatives (options or choices, 
that we enumerate in an arbitrary order), 
as well as an ``undecided'' state $s_0$, assessed when the individual has not adopted a defined option.  
The addition of the undecided state reflects the fact that sometimes people do not have  a favorite choice.

 We focus on the dynamics developed in   ER and BA networks, 
as representative of small-world networks with homogeneous and heterogeneous degree distributions, respectively. 
However, for comparative purposes we will also consider random neighbors and nearest neighbors in a square lattice 
(with periodic boundary conditions). 

We  assume  that most individuals do  not have a formed opinion {\it a priori}, 
except for initiators representative of each offered choice. 
Then, we start the dynamics with all nodes in the $S=s_0$ state,  
except randomly chosen $q$ nodes, to each of which we attribute a different opinion $S=s_1, \ldots, s_q$.
We consider the same number of initiators  (one initiator)  for each alternative, 
reflecting the equivalent attractiveness of all the alternatives.
This kind of initial state  has been  used in   opinion models for proportional elections~\cite{Travieso2006a}.

At each  Monte Carlo (MC) step of the dynamics, we visit all the nodes of the network  in a random order and 
update them successively, in asynchronous mode. 
The state of the visited node $i$ is updated  according to the following steps: 

\begin{enumerate}[(i)]
	\item We define the set of nodes, ${\cal A}_i$, formed by  $i$ and its nearest neighbors.  
	\item We determine the plurality state $S^p_i\neq s_0$, associated to node $i$, 
	as the state shared by the largest number of nodes in the set ${\cal A}_i$.
	\item The agent $i$ will then adopt its corresponding plurality state.
\end{enumerate}

\begin{figure}[b!]
\begin{center}
    \includegraphics[width=0.75\columnwidth]{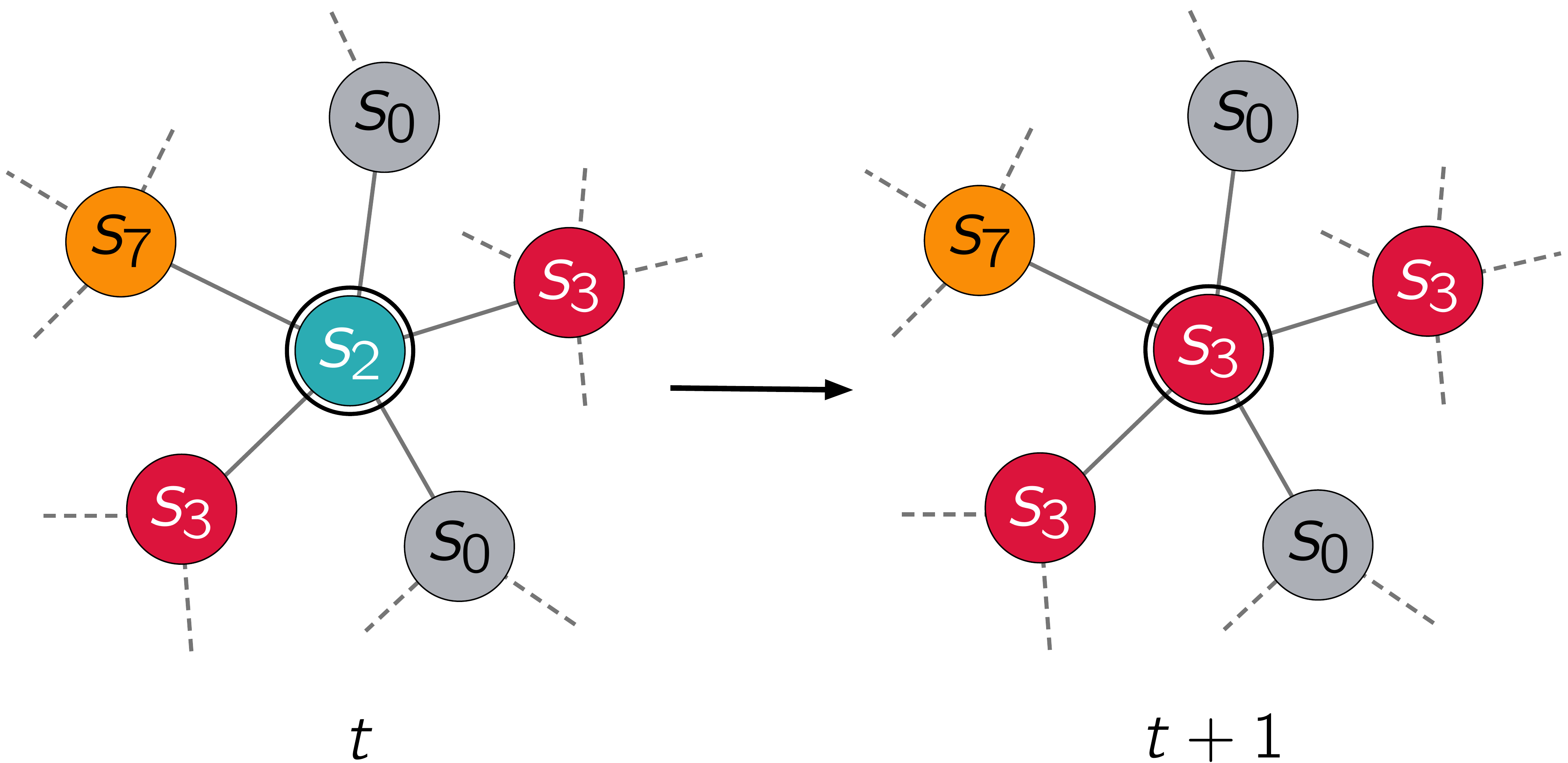}
\end{center}
\caption{Illustration of the model rule. 
In a given instant $t$, a node $i$ (the central one in the figure) and its neighbors form a group ${\cal A}_i$. 
In the case shown in the figure, the plurality state is $S^p_i= s_3$, 
since this is the state which is shared by more nodes. 
So, in the next time step, the node $i$ changes its opinion from $s_2$ to $s_3$. 
Notice that nodes with $S=s_0$ are ignored in the computation of the plurality state $S^p_i$.} 
\label{fig:rules} 
\end{figure}

Notice that, when we measure the state $S^p_i$, we ignore the nodes in ${\cal A}_i$ that have $S=s_0$ 
(see Fig.~\ref{fig:rules}) but 
the current opinion of site $i$ also counts to define $S^p_i$.

The updates are repeated  and the dynamics stops when an absorbing state is attained, i.e., 
if, at a MC step, none of the nodes changes its state.

Let us remark  that this dynamics differs from that of the Sznajd type~\cite{Sznajd-Weron2001a,Slanina2003a} where two or more individuals
sharing the same opinion impose it to all their neighbors. 
It also differs in several aspects from the plurality rule introduced in Ref.~\cite{Chen2005a}: 
i) while in our case  only the central node is affected, in~\cite{Chen2005a} the whole group ${\cal A}_i$ changes its opinion to
 the plurality state $S^p_i$ in a single update step,  
ii) here the size of the interaction group ${\cal A}_i$ is given by $G=k_i+1$, where  $k_i$ is the connectivity of site $i$, 
instead of being constant   
(anyway, the parameter $\langle k \rangle+1$ plays the role of an effective $G$, and they coincide in the limit of a highly homogeneous, 
or  regular,  network); 
iii)  the possibility of indecision is not contemplated in~\cite{Chen2005a};   
iv) in an event of tie, the opinion of node $i$ remains unchanged in our model, 
while,  in~\cite{Chen2005a}, one of the dominant options is randomly selected.  
With respect to this last item, the present dynamics is more close  to the majority rule version of Ref.~\cite{Chen2005a}, 
where the dynamics becomes static because, when there is no local majority, the state of the group does not change.
v) In terms of the underlying network, here we consider small-world networks, while the dynamics in Ref.~\cite{Chen2005a} 
was studied in the mean-field limit and over a square lattice. 
vi) Finally, another important difference is in the initial conditions: we consider that decided nodes are diluted in a sea of undecided nodes, 
 instead of equiprobability of definite opinions.

\section{Results}
\label{sec:results}

\subsection{Plurality dynamics}
\label{sec:dynamics}

\begin{figure}[h!]
\includegraphics[scale=0.82]{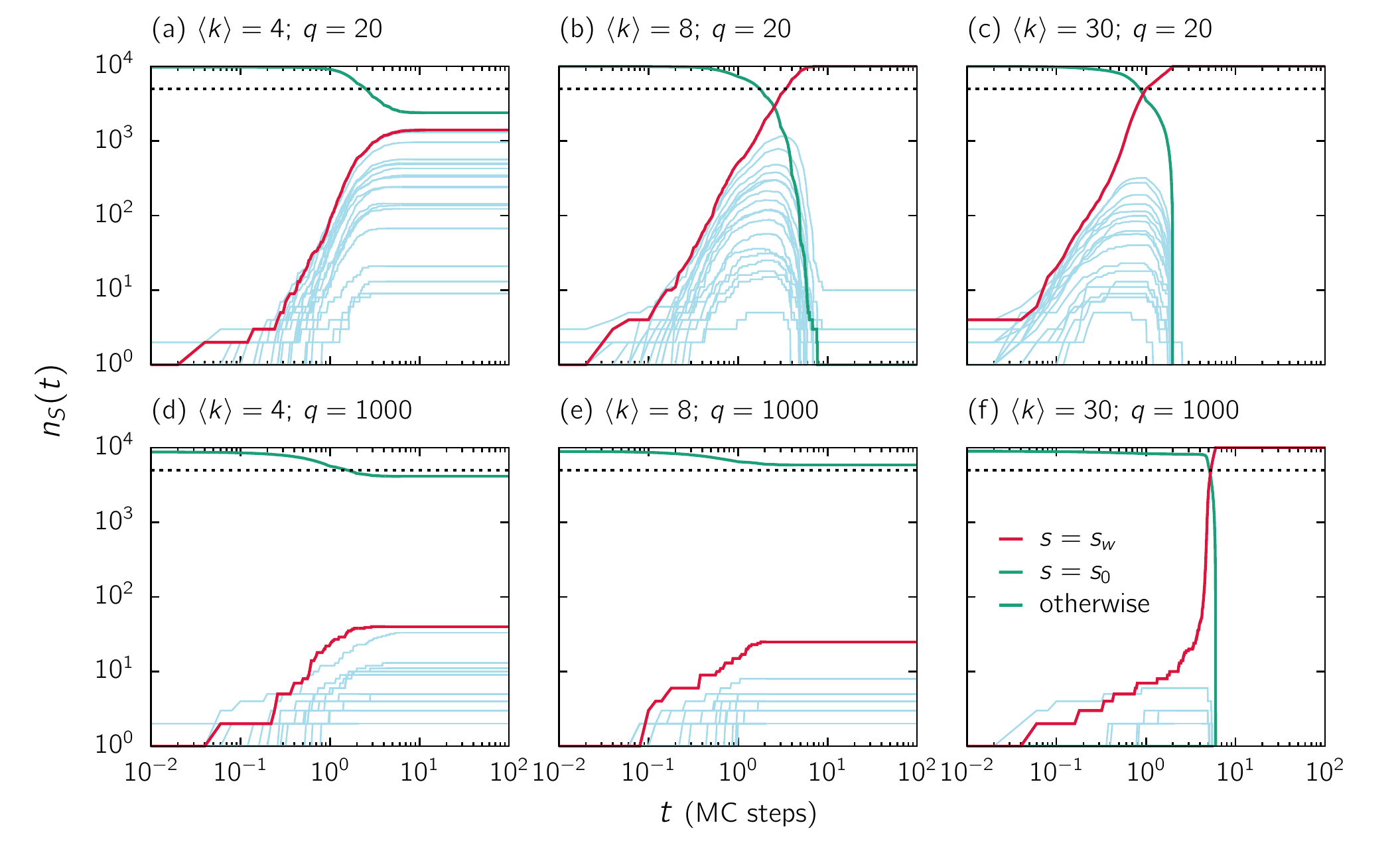}
  \caption{Time evolution of $n_s$, the number of nodes that share the same opinion $s$, 
	for an ER network of size $N=10^4$.  For $q=20$, all opinions are shown, while for $q=1000$, only 
	 the winner opinion and other 20 opinions are shown.  
The winner opinion $S=s_w$, as well as the opinion $S=s_0$ (undecided state)  are highlighted. 
The level 50\% of the population  is represented by black dotted lines.  
	The stationary state is attained in a few MC steps and  distinct patterns arise 
	for different values of $\langle k\rangle$ and $q$.  
}
  \label{fig:dynamics}
\end{figure}

We follow the evolution of each realization of the dynamics until the final state is attained. 
Distinct distributions of opinions can emerge in the final state depending 
on the amount of alternatives $q$, the average connectivity $\langle k \rangle$,  the 
network topology and  size $N$. 
Fig.~\ref{fig:dynamics} illustrates the evolution of $n_s$, the number of nodes that share the same opinion $s$, 
in representative realizations of the dynamics on ER networks  of size $N=10^4$. 
A final state is reached in a few Monte-Carlo iterations. 

Fig.~\ref{fig:dynamics} illustrates the distinct patterns that can arise, 
while the full phenomenology as a function of the model parameter $q$ 
and network features will be shown in Sec.~\ref{sec:diagram}.

\noindent
(a) For instance, when $q=20$ and $\langle k\rangle=4$, we have a fragmented final state, 
where all initial values of $S$ 	have adopters. 
The number of undecided agents is predominant, representing about 20\% of the population, in this particular realization. 

\noindent
(b) For larger connectivity (e.g., $\langle k\rangle=8$), 
	several opinions are  still possible in the stationary state, 
	but the winner opinion $s_w$ is widely dominant. 
	The number of undecided agents in the steady state has decreased significantly with respect to case (a).

\noindent 
(c) For even larger connectivity (e.g., $\langle k\rangle=30$),   consensus is likely. 

\noindent
(d)-(f) When $q$ is large, the number of undecided agents does not decrease monotonously when the 
connectivity increases. But, increasing the  connectivity, consensus is reached, although it can take 
a larger time than for small $q$.  
  
In all cases, the number of undecided nodes decreases with time, because  undecided individuals are not  produced 
by the dynamics in the present version of the model, but  only introduced in the initial condition.

The dynamics in ER networks can be qualitatively understood as follows:
 
In a first regime, each  opinion propagates invading the undecided neighbors. 
If the initiators are very diluted ($q \ll N$), and the connectivity is not too high, 
then each cluster of nodes with the same opinion  can develop  almost independently of each other, during several MC steps 
(non-competitive regime). 
In this case, the initial growth is nearly exponential,  described approximately by $dn_s/dt=\langle k\rangle n_s$ in ER networks.
  
When two or more clusters collide, a competitive regime starts.   
Depending on the network, the competition can take place more or less evenly so that ties stagnate the evolution 
avoiding wide dominance of a given opinion (as in Fig.~\ref{fig:dynamics}.a, 
\ref{fig:dynamics}.d and \ref{fig:dynamics}.e). 
Otherwise,  a sort of rich-get-richer or cumulative advantage mechanism can take place. 
In that case, the winner opinion becomes noticeably larger than other ones,  
 convincing individuals from other opinion clusters  
(Fig.~\ref{fig:dynamics}.b) or even the whole network (Fig.~\ref{fig:dynamics}.c). 

\begin{figure}[h!]
\includegraphics[scale=0.82]{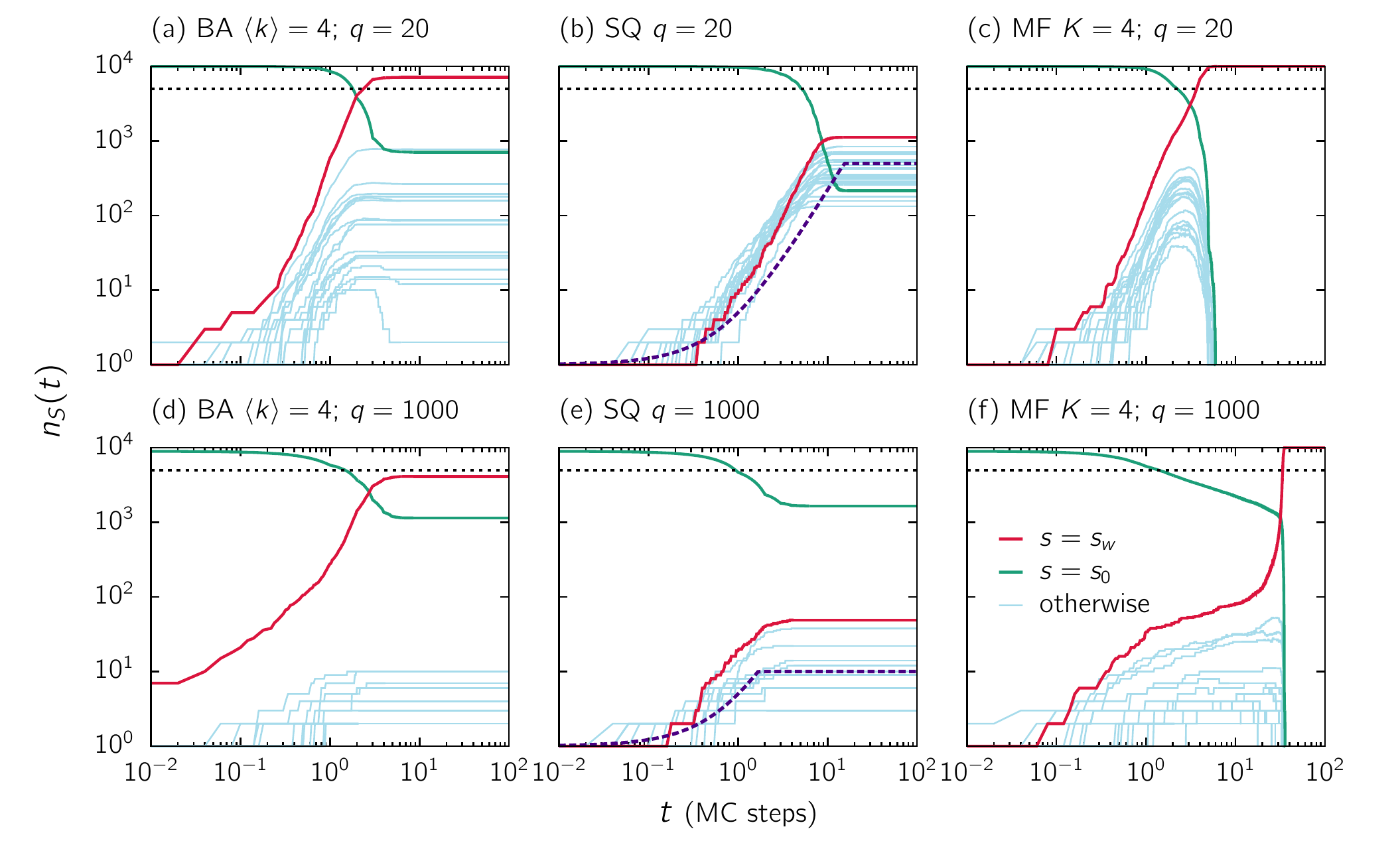}
  \caption{Time evolution of $n_s$,  with the same conventions of  Fig.~\ref{fig:dynamics}, 
	for BA networks, square lattices (SQ) and mean-field (MF) interactions (random networks) for the parameters indicated 
	in the figure. In panels (b) and (e) the purple dashed lines are given by Eq.~\ref{sq}. 
	In all cases $N=10^4$.
}
  \label{fig:dynamics2}
\end{figure}

Similar patterns as those shown for ER networks are also observed for BA networks, although for different values of the parameters, 
as illustrated in the first column of Fig.~\ref{fig:dynamics2}.
Notice that the winner opinion, as well as the number of decided people, for 
the same parameters,  are favored in BA networks, where cumulative advantage effects are more accentuated.

Let us remark that in some extreme situations, given the initial conditions studied here (one initiator for each state), 
the system does not evolve.  
This occurs, for instance, in the limit case when  $q=N$ (hence, each individual has a defined opinion) 
or   the connectivity of all sites is $N-1$ (complete graph). In those extreme cases,  ties forbid changes of state and 
the dynamics is frozen from the start.  
However, we will restrict the range of the parameters to the region  $q, \langle k \rangle \ll  N$.

Aside from networks with the  small-world property (ER and BA),  we also analyzed, for comparison,   interactions   with 
(i)  first neighbors in square (SQ) lattices (with periodic boundary conditions), and 
(ii) $K$ random neighbors (with $K>1$). 
Representative examples of the evolution, in square lattices  and $K$ random neighbors  
for the case  $K=4$, are shown in the second and third columns of Fig.~\ref{fig:dynamics2}.

In regular lattices, the evolution   is essentially non-competitive. 
Clusters grow from their initiators and, when they collide, 
the dynamics freezes due to ties in the interfaces, without entering a competitive phase,  
in contrast to  ER and BA networks where there are long-range links that break ties. 
Since  the initial growth occurs at the surface of the cluster, then $dn_s \propto \sqrt{n_s}$ 
which, differently from small-world networks, 
gives a quadratic increase of $n_s$ with $t$, as observed in  Fig.~\ref{fig:dynamics2} (second column). 
Alternatively, it is easy to show that, for the synchronized update, at short times, 
the number of adopters of each choice grows around its initiator,  following, in average, the recursion relation 
$n_s(t)= n_s(t-1)+ Kt$, where $K=4$ for the square lattice. 
By solving this recursion equation, one obtains 
\begin{equation} \label{sq}
n_s(t)=n_s(0)[1+Kt(t+1)/2],
\end{equation} 
that yields  the  predicted quadratic increase with $t$, valid for  small $t$,  until   
$n_s\simeq N/q$ holds.  Despite the prediction is done for the synchronous update, it is in good agreement with the 
average value of the simulated curves, as can be seen in the second column of Fig.~\ref{fig:dynamics2}.

As a consequence of the lack of competition, the final values of $n_s$ are less disperse in the square lattice 
than in random networks and, mainly, consensus, or even a wide dominance of an opinion, becomes unlikely for $q>1$.

In the absence of any network structure (i.e., when neighbors are chosen purely randomly), like in the examples of the last 
column of Fig.~\ref{fig:dynamics2}, one of the opinions dominates and attains consensus, even for very small $K$. 
This can be understood in terms of a mean-field approach,  following the lines of Ref.~\cite{Chen2005a}. 
In fact, the fraction of undecided sites $f_0= n_{s_0}/N$ follows an equation of the type
\begin{equation} \label{zero}
\dot f_0 = -f_0 P_{K-1}(f_0)\,,
\end{equation}
where $P_{K-1}$ is a polynomial of order $K-1$ in $f_0$ whose coefficients depend 
on the fractions $f_s\equiv n_S/N$, with $S=s_1,\ldots,s_q$ and $P_{K-1}(0)>0$. 
For all $K$ and $q$, the factor $f_0$ arises from the central undecided node that is part of the group. 
Therefore, in the steady state it must be $f_0=0$. Moreover,  since $P_{K-1}(0)>0$,  Eq.~(\ref{zero}) is stable for $f_0=0$.
The remaining equations for  the other fractions $f_s$ have solutions of the type found in Ref.~\cite{Chen2005a}  
for their majority and plurality versions. 
In particular, the stable solutions are those of consensus, 
where $f_{s_j}$=1 for  some $j$ (hence, the remaining fractions vanish).  
 
For instance, for $K=2$ and $q=2$ (let us call  $f_{s_j}\equiv f_j$, $0\le j\le q$), we have
\begin{eqnarray}
\dot f_0 &=& -f_0[f_1^2+f_2^2+2(f_1+f_2)f_0 ]\,, \nonumber \\
\dot f_1 &=&  f_0f_1(2f_0 + f_1) +f_1f_2(f_1-f_2) \,, \nonumber \\     
\dot f_2 &=&  f_0f_2(2f_0 + f_2) +f_1f_2(f_2-f_1) \,. \nonumber
\end{eqnarray}
It is easy to obtain that the only stable solutions are $(f_0,f_1,f_2)= (0,0,1)$ and $(f_0,f_1,f_2)= (0,1,0)$.

Similarly, for $K=3$ and $q=2$, once $f_0=0$, we have
\begin{eqnarray}
 \dot f_1 &=&   f_1f_2(f_1-f_2)(f_1+f_2) \,, \nonumber \\
 \dot f_2 &=&   f_1f_2(f_2-f_1)(f_1+f_2) \,, \nonumber
\end{eqnarray}
that lead to the consensus solutions as the stable ones,  
while equipartition $f_1=f_2=1/2$ is unstable. 
The  equations above, for small values of $K$ and $q$,  valid for $f_0=0$ in the present model, 
are the same obtained for the majority and plurality versions studied in Ref.~\cite{Chen2005a}, 
although the equations in three cases differ for enough large  values of $K$ and $q$.

Increasing $K$ leads to equations of the form $\dot f_1 = f_1f_2(f_1-f_2)P_{K-2}(f_1)$, 
where $P_{K-2}$ is a definite positive polynomial of order $K-2$ in $f_1$, 
whose coefficients depend on $f_2$. 
Therefore, consensus is always stable for $q=2$.

For large number of alternatives, $f_0$ necessarily must vanish as well, and consensus is also a stable solution. 
For instance when $q=3$ and $K=2$, once $f_0=0$, we have
\begin{eqnarray}
\dot f_1 &=&   f_1^2(f_2+f_3)-f_1(f_2^2+f_3^2)  \,, \nonumber \\
\dot f_2 &=&   f_2^2(f_3+f_1)-f_2(f_3^2+f_1^2)   \,, \nonumber\\
\dot f_3 &=&   f_3^2(f_1+f_2)-f_3(f_1^2+f_2^2)   \,, \nonumber
\end{eqnarray}
that lead to the consensus solutions as the stable ones, 
while equipartition $f_1=f_2=f_3=1/3$ is unstable, and the solutions of the 
type $(f_1,f_2,f_3)=(0,1/2,1/2)$ are saddle points.
 
Increasing $q$ and $K$, the structure of fixed points becomes more complex and 
more routes to consensus emerge~\cite{Chen2005a}. 
Nonetheless,  consensus is always the final state, which was also verified through numerical simulations.

Differently, when the structure of the interaction network is relevant, non-trivial behaviors occur, as those illustrated 
in Fig.~\ref{fig:dynamics}.  
A population of undecided people can survive and consensus is not always attained.

\subsection{Phase diagram}
\label{sec:diagram}

To summarize, the nontrivial final configurations that emerge in ER and BA networks, we built  a phase diagram 
in the plane $\langle k\rangle - q$.   
For each realization, we monitored the fraction of decided people 
\begin{equation} 
f_{d}\equiv n_d/N  \equiv (N-n_{s_0})/N  \,,
\end{equation}   
and also the fraction of nodes sharing  the most adopted opinion, or winner choice, 
\begin{equation} 
f_{w}\equiv n_{s_w}/N \,.
\end{equation}  
 
These quantities were averaged at the final state of several realizations. 
Unless said something different, at least  50 realizations were considered for each set of values of the parameters.
For each realization of the dynamics, a different network was generated.  
The averaged fractions will be denoted by  $\overline{f_{d}}$ and $\overline{f_{w}}$, respectively.  
 
We  also computed the fraction  $p_{c}$ of the simulations that reach  consensus 
(operationally meaning at least 99\% of the population).

 When performing computations over ER networks, only the main component of the graph was considered.

\begin{figure}[ht!]
\includegraphics[scale=0.82]{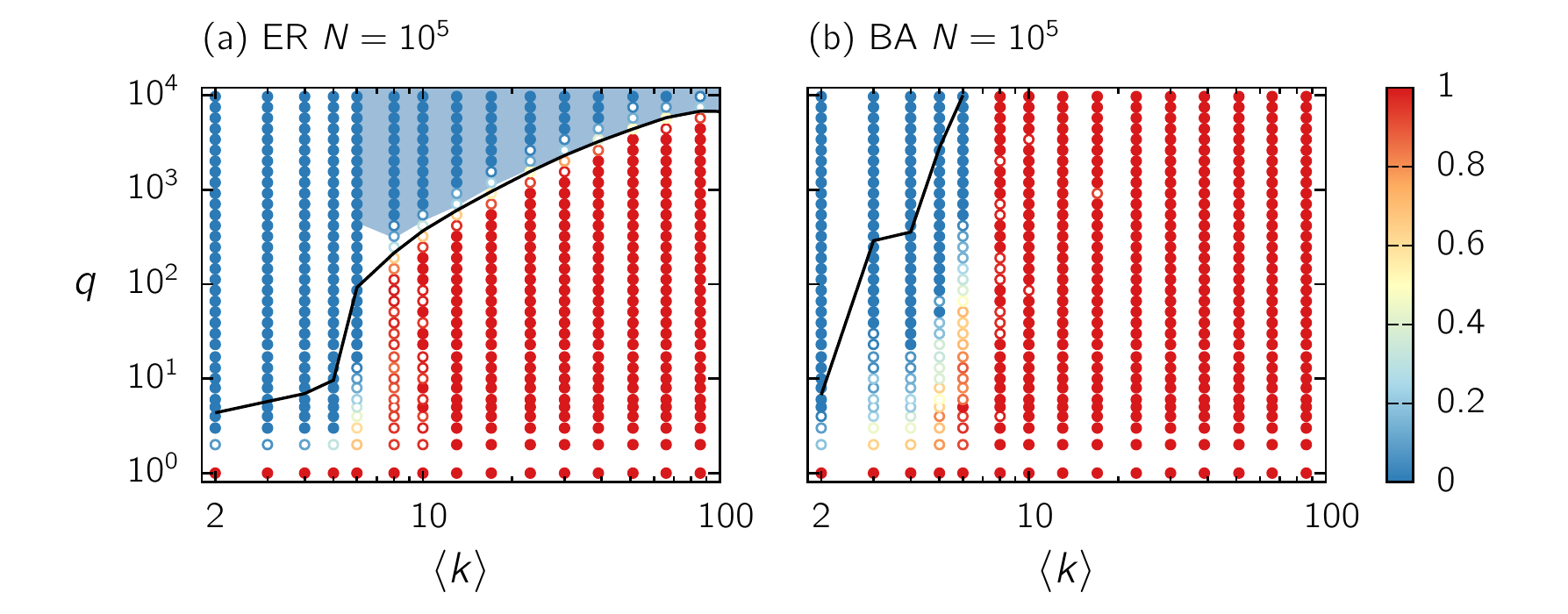}
  \caption{
    Phase diagram in the plane $\langle k\rangle-q$ showing the fraction of consensus $p_c$ in color scale. 
	Full symbols emphasize the points where consensus occurs in all  ($p_c=1$, dark red) and none  ($p_c=0$, dark blue) of the 30 realizations for each cell, 
		otherwise symbols are empty.  
		The solid lines depict the points such that $f_w=0.5$ ($f_w>0.5$ to the right of the curves). 
		The shadowed areas represent the points for which $f_d<0.5$.
    ER and BA networks size is  $N=10^5$.
}
  \label{fig:diagram}
\end{figure}

The phase diagram in the plane of parameters $\langle k \rangle-q$, for ER and BA networks is depicted in 
  Fig.~\ref{fig:diagram}.  We restricted the analysis to the  region   $q\le N/10$ and $\langle k \rangle \le N/1000$.
The diagram shows the changeover from regions of consensus to regions of fragmented final state, 
as indicated by the colors from red to blue.  
Filled symbols emphasize the points where consensus is certain ($p_c = 1$, red) or uncertain ($p_c = 0$, blue).
Moreover, the solid lines depict the frontier where $f_w=0.5$ ($f_w>0.5$ to the right of the curves), 
and the shadowed areas highlight the points for which undecided people are majority ($f_d<0.5$). 
It is clear that the  consensus domain ($p_c\simeq 1$, red region) is larger for BA networks, 
pointing out that heterogeneity favors consensus. 
In both ER and BA networks, inserting a few links, for fixed $q$ near the transition frontier, may trigger consensus.

 In BA networks, for sufficiently large $q$ ($\gtrsim 200$) 
the critical value   $\langle k \rangle_c \simeq 7$ becomes independent of $q$. 
In contrast, in ER networks, the dependence on $q$ is stronger. 
The non-consensus domain for ER networks, ($p_c\simeq 0$, blue region), when  $q$ becomes sufficiently large, 
spreads over the region of large mean connectivities. 
It means, that near the transition frontier, eliminating a few alternatives can trigger consensus, 
an effect which in BA networks only occurs for  connectivities below $\langle k \rangle_c \simeq 7$ .

Concomitantly, in the shadowed area in Fig.~\ref{fig:diagram}.a.,  
the fraction of decided people becomes minority ($f_d<0.5$). 
Differently, in the BA case, the majority of nodes is decided  over the whole phase diagram.   
Moreover, even in the absence of consensus, the winner group can become majority ($f_w>0.5$), more easily  in BA networks. 

In the following sections \ref{sec:q} and \ref{sec:k}, 
we describe in more detail the dependency on $q$ and $\langle k \rangle$, respectively.

\subsection{Effect of the number of options $q$}
\label{sec:q}

In this section, we  focus on the impact of the number of opinions $q$ on the steady state, and we also discuss size effects. 

\begin{figure}
\includegraphics[scale=0.82]{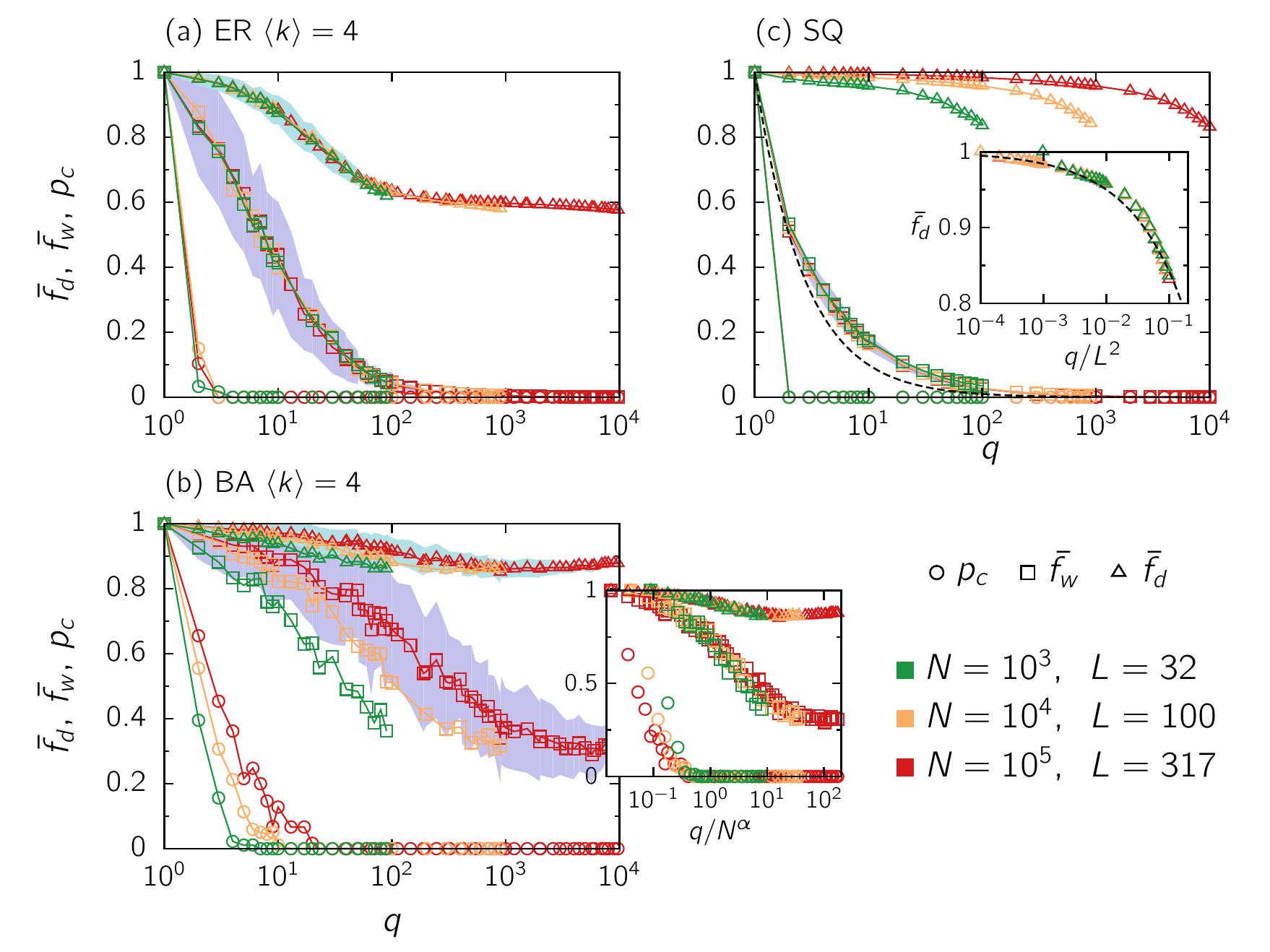}
  \caption{
Average fraction  of decided people $\overline{ f_d }$ (triangles), 
average fraction of the winner choice $\overline{ f_w }$ (squares),
and fraction of realizations reaching consensus  $p_c$ (circles),  
as a function of the 	number of different opinions $q$. 
Three different sizes were used for ER and BA  networks with mean connectivity $\langle k \rangle =4$, and for square lattices. 
The number of nodes $N$ is given for networks and the linear size $L$ for lattices. 
	Solid lines are a guide to the eye, and  shadowed regions represent  
	the standard deviations of $f_d$ and $f_w$ in the case $N=10^5$ 
	(that are similar for the other sizes).  In panel (c), the dashed line corresponds to $1/q$. 
		Computations were done over at least 50 samples. 
The insets  show   the data collapse, as explained in the text: 
in (b), $\alpha=0.35$; in (c),  the dashed line is given by Eq.~(\ref{fdSQ}). 
}
  \label{fig:opinions}
\end{figure}

Fig.~\ref{fig:opinions} shows  the three quantities of interest, 
$\overline{ f_d }$,  $\overline{ f_w }$ and   $p_c$,  as a function of $q$, for ER and BA networks and SQ lattices  
of three different sizes. 
The plots of $p_c$ vs $q$, for ER and BA networks, correspond  to vertical cuts of the phase diagram in 
Fig.~\ref{fig:diagram}. 
All the fractions monotonically decrease with $q$, in the studied range, 
indicating that independently of the lattice, increasing the number of options hinder the process of decision making, 
promoting indecision, and also turning more difficult the appearance of a popular choice.  

Concerning size effects, in ER networks (panel a),  the three quantities are almost independent of size $N$. 

The results in  square lattices (panel c)  are in accord with the predictions made 
in Sec.~\ref{sec:dynamics}. For instance,  $p_c>0$ only for $q=1$. 
The winner fraction $\overline{ f_w }$ is independent of $L$ and decays subtly above $1/q$, as expected. 
In fact,   we have seen that the fractions of each opinion have a narrow distribution around the mean value $1/q$. 
Meanwhile, the decided fraction ${ f_d }$ depends on $L$. 
Since undecided sites are at the interfaces of each opinion cluster, 
then, their quantity  over all $q$ clusters is $n_{s_0}\propto  \sqrt{L^2/q} \times q/2$. 
Therefore,  the decided fraction $f_d=1-n_{s_0}/L^2$ results
\begin{equation} \label{fdSQ}
 f_d= 1-\sqrt{q/L^2}/2 \,.
\end{equation} 
As a consequence, the curves of   $\overline{ f_d }$  for different sizes collapse when represented vs 
$q/L^2$, as shown in the inset of Fig.~\ref{fig:opinions}.c.   
 
In contrast, in BA networks (panel b), $p_c$ and $\overline{f}_w$ also depend on $N$.  
Moreover,  $\overline{f_w}$ does not vanish but tends to a finite value as $q$ increases. 
The heterogeneity of these networks favors that the winner  conquers a large domain, 
which increases with system size. 
The decided fraction $\overline{ f_d }$ also depends on $N$ and data 
approximately collapse  when represented vs  $q/N^\alpha$, with $\alpha \simeq 0.35\pm 0.05$ for the case of Fig.~\ref{fig:opinions}.b (see inset). 
This value of $\alpha$ suggests that  the  number of ``interfacial'' nodes     
scales with the number of bulk nodes in a cluster  with exponent $1-\alpha\simeq 0.65$. 
The fraction $\overline{ f_w }$, as well as the critical value at which $p_c$ vanishes,  follow the same scaling, 
as shown in the inset of Fig.~\ref{fig:opinions}.b. 

The average fraction of decided people $\overline{f_d}$   typically  
decreases with $q$ (for $q \ll N$), but a kind of saturation effect occurs for large enough $q$ 
and a flat level appears in small-world networks, 
indicating that $\overline{f_d}$   becomes insensitive to the 
introduction of new choices. However, the value of the flat level changes   
with the network connectivity (not shown).  
For large number of options $q$, the values of $\overline{f_d}$ are smaller in ER networks.
Heterogeneity of degrees seems to be helpful in breaking ties. 
In the heterogeneous BA networks, many nodes have low connectivity and can be easily 
convinced by a decided neighbor. 
On the other hand, the occurrence of  a local plurality is less likely 
in an homogeneous ER network with a given 
connectivity. As a consequence, ties  are more frequent,  more nodes remain undecided and 
the dynamics freezes. 
However, in square lattices, despite the homogeneity,  the undecided fraction is relatively small. 
This can be understood as follows. Ties occur when distinct opinion clusters collide,  then, 
the surviving undecided nodes are located at the ``interfaces''. 
In networks with long-range links the encounter of different clusters occurs early, and many nodes remain undecided, 
meanwhile in regular lattices with nearest-neighbor interactions, 
undecided nodes are being conquered until the collision,   late in the dynamics,  
when the opinion groups have occupied most of the lattice and few undecided nodes remain 
at the interfaces.

The average fraction of the population adopting the winner option, $\overline{f_w}$ (squares) is also a significant 
quantity. (Necessarily $\overline{f_w} \le \overline{f_d}$.) 
The fraction $\overline{f_w}$  is greater in BA networks. 
That is, the winner choice conquers in average a large  fraction of the population in  BA networks, compared to ER networks 
and square lattices with equivalent  $\langle k\rangle$.
In fact, the cumulative advantage  that drives the growth of an opinion group
is facilitated  in these  heterogeneous networks due to the presence of hubs, 
and the winner conquers more adopters. 
Also notice that, in ER networks, when the fraction $f_d$ attains the flat level, 
 a dominant opinion is absent, as mirrored by the  very small value of $\overline{f_w}$ 
(see Fig.~\ref{fig:opinions}.a).
Meanwhile, in BA networks the winner can always conquer  an important fraction of the population 
(Fig.~\ref{fig:opinions}.b), shown by the fact that  $\overline{f_w}$ remains finite 
(except in the  limit $q\to N$).

For all kinds of networks, the probability of occurrence of consensus, 
$p_c$ (circles), typically falls  from 1 to 0 as $q$ increases. 
This agrees with the intuition that, when there are more options to choose, 
it is more difficult to attain consensus. 
The probability of consensus decays rapidly with $q$ and,  above a critical value $q_c$, 
the fraction $p_c$ becomes negligibly small.  
This effect is accentuated in the square lattice where for $q_c=1$.

\subsection{Effect of the mean connectivity $\langle k \rangle$}
\label{sec:k}

\begin{figure}
\includegraphics[scale=0.82]{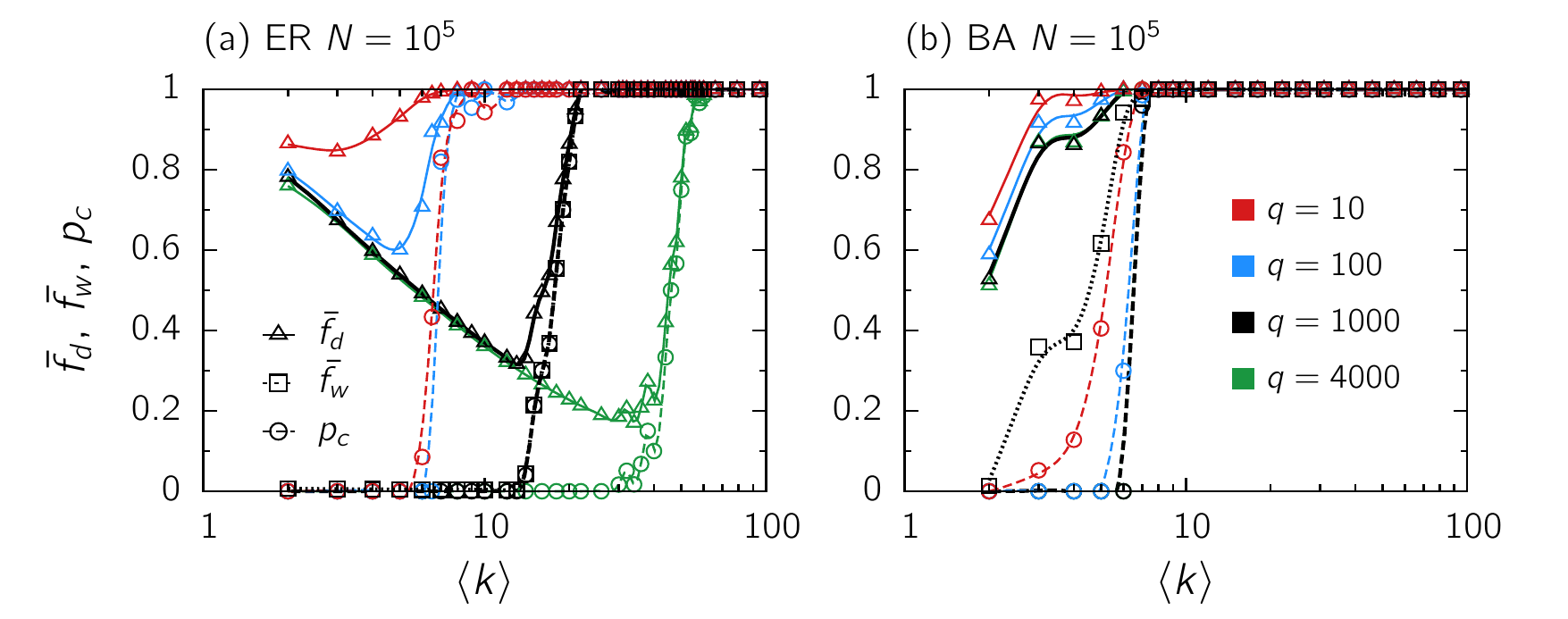}
  \caption{
Fractions $\overline{f_d}$ (full lines) and $p_c$ (dotted lines) as a function of the mean connectivity  $\langle k \rangle$, for several values of $q$, indicated by different colors, in ER (a) and BA (b) networks of size  $N=10^5$. 
For the sake of clearness, the behavior of the winner fraction  $\overline{ f_w }$ (black dotted lines) is depicted only for the case $q=1000$. 
Averages where computed over  at least 50  samples.    	
}
  \label{fig:connectivity}
\end{figure}

The behavior of the characteristic fractions $\overline{f_d}$, $\overline{f_w}$  and $p_c$   as a function 
of the average connectivity of the network, $\langle k \rangle$, are  shown in Fig.~\ref{fig:connectivity}, 
for several values of $q$, in ER and BA networks. 
 The plots of $p_c$ vs $\langle k \rangle$  correspond  to horizontal cuts of the phase diagram in 
Fig.~\ref{fig:diagram}. 
 
Let us start by the case of  BA networks (shown in Fig.~\ref{fig:connectivity}.b) that exhibits a simple monotonic behavior, for the range  shown in the figure. 
The three fractions increase  with the connectivity. 
As observed   in the phase diagram of Fig.~\ref{fig:diagram}.b, Fig.~\ref{fig:connectivity}.b   shows in more detail 
how the jump to consensus becomes more abrupt as $q$ increases and the critical value of the connectivity 
becomes nearly independent of the number of options ($\langle k \rangle_c \simeq 7$), as also observed in 
the phase diagram.  

Differently, in ER networks (see Fig.~\ref{fig:connectivity}.a), $\langle k \rangle_c$ increases with $q$. 
Moreover,   the average fraction of decided people $\overline{f_d}$ first   
 decreases with the connectivity down to a minimal  value localized  at $\langle k \rangle_{\rm min}$. 
Up to that point, the fraction of simulations attaining consensus $p_c$ is  negligibly  small. 
But, at $\langle k \rangle_{\rm min}$, a transition occurs and 
both $p_c$ and  $\overline{f_d}$  rapidly increase  with $\langle k \rangle$, up to 1. 
One would expect to have more decided nodes when the connectivity is higher, like in the case of BA networks, 
since, in principle, more connections might facilitate  information spreading. 
However, on a low connected network, opinion groups are typically isolated from each other. 
When links are added, and disconnected groups become connected,   ties can occur. 
That is, on the one hand higher connectivity implies that groups of different opinions 
can be more connected among them and compete. On the other, a node will be aware of more opinions, 
making difficult the decision and keeping more undecided nodes. 
Therefore, not only overchoice (high $q$)  may produce stagnation of the dynamics but also ``overlink'' 
or excess of contacts due to high $\langle k \rangle$. 
This explains  the initial decrease of $\overline{f_d}$ with $\langle k\rangle$ which occurs up to a minimal value 
of $\overline{f_d}$. After that point, introducing more connections will allow a dominant group to impose its opinion, 
concomitantly $p_c$ increases until reaching its maximal value 1.
Also notice that in ER networks,  for large values of $q$, there is an interval of mean connectivity for 
which the fraction of decided people becomes minority.

The existence of an abrupt transition from a situation  where many opinions 
coexist to consensus indicates that,  by adding just a few  links or by removing a few  choices, 
most people may come to adopt the same state. 
The transition to consensus is more abrupt in BA networks, and the jump width decreases with $q$. 
In these networks, as discussed above, there is a dominant winner opinion group,  
that represents an important fraction of the population (finite $f_w$). 
The largest group  gains additional adopters more easily, with cumulative advantage.
Near the critical connectivity,  when adding few links at random, 
it would be more probable to connect the very large group to smaller ones, and, 
as a consequence, they would be conquered by the dominant opinion, rapidly leading to consensus. 
In homogeneous ER networks this transition is less abrupt, because a  largely dominant group is less probable.  
The width of the transition region slightly increases with $q$. 
For the extreme case of nearest neighbors interactions in square lattices, cumulative effects are completely absent, therefore, a transition to consensus is unlikely.

\subsection{Distribution of opinions and empirical data}
\label{sec:pdfs}

\begin{figure}
\includegraphics[scale=0.82]{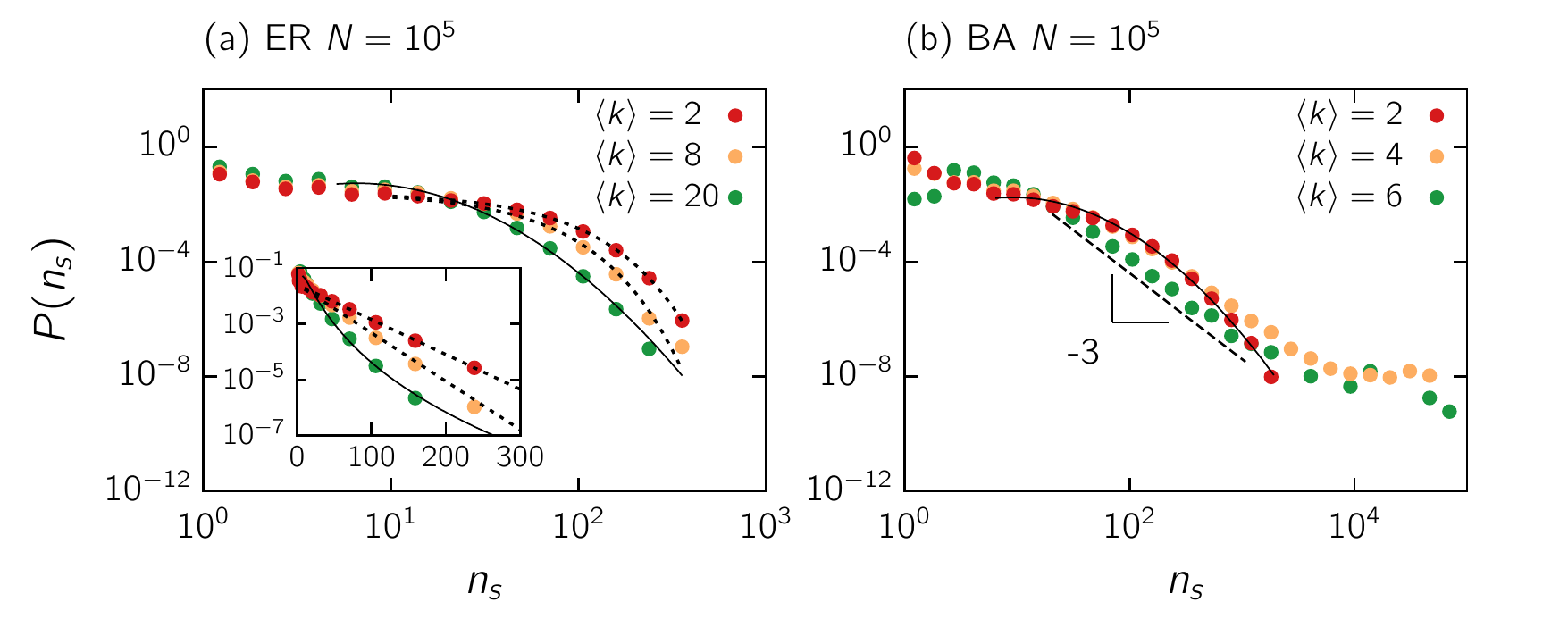}
  \caption{
    Distribution of the number of nodes that adopted a given opinion  $n_s$, $P(n_s)$, 
		built over 50 samples, for $q=2000$, on ER  (a) and BA (b) networks with size $N=10^5$, 
			and different values of  $\langle k \rangle$ indicated on the figure.    
	  The solid lines are log-normal fits, the dotted lines exponential fits, and the dashed line 
		with slope -3 was drawn for comparison. }
  \label{fig:pdfs_k}
\end{figure}

In non-consensus steady states,  a broad distribution of opinions across the population  can emerge.   
In order to analyze the shape of the distributions, 
we built the normalized histograms of $P(n_s)$, where $n_s$ is the number of nodes with a given opinion $s$. 
Histograms were computed by accumulating realizations ending in non-consensus states.
Typical  distributions are depicted in Fig.~\ref{fig:pdfs_k}. 
One can identify exponential, log-normal and power-law behaviors.

In the ER case, far enough from the critical frontier of consensus, the preferences are almost uniformly distributed 
with an exponential cutoff. 
When approaching consensus (for instance by increasing $\langle k\rangle$) 
the distribution adopts a  log-normal shape.  Notice that this occurs in the region of the phase 
diagram where indecision prevails (shadowed area in Fig.~\ref{fig:diagram}).
In BA networks, the distribution can also resemble  a log-normal, but when approaching consensus the tail rises due to the existence of dominant winners.
Moreover, when the dynamics freezes early, $P(n_s)$ tends to reflect the  degree distribution with exponent $-3$.

\begin{figure}
\includegraphics[scale=0.82]{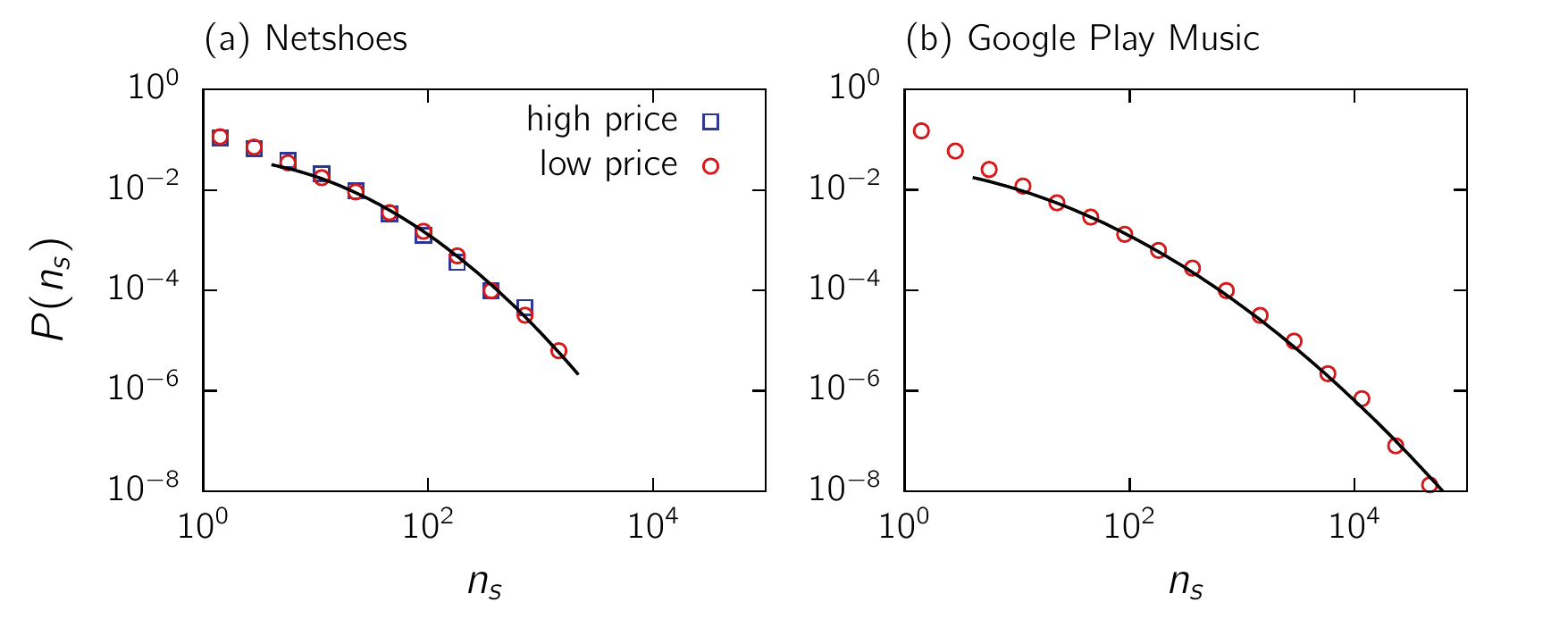}
  \caption{
    Distribution $P(n_s)$, of the number of (favorable) reviews $n_s$, with 4 and 5 stars, 
		for  Netshoes~\cite{netshoes}   and   		Google Play music~\cite{googlemusic}. 
		It is reasonable to identify  the number of items with $q$, and 
		the number of reviews giving 4 and 5 stars with the number of decided people $n_d$. 
		For Google Play musics, prices are similar (U\$ $9 \pm 3$), then all data were used 
			($n_d\simeq 1.6\times 10^6$; $q\simeq 7500$). 
		For Netshows we split the data into two subsets: items 		with prices 
		above ($n_d\simeq 9\times10^4$; $q\simeq 1720$) and below ($n_d\simeq 10^5$; $q\simeq 1720$) the median ($\simeq$ R\$  200). 
		The solid lines are log-normal fits.
	  }
  \label{fig:pdfs-real}
\end{figure}

In order to compare the distributions from simulations with those from real world, 
  we considered products 	than are rated online. 
	We analyzed data about items whose alternatives are not significantly differentiated 
	(for 	instance, in price and/or quality), as assumed in 	our model. 
	We identified $q$ with the number of items within each category  
	and we considered the amount of positive reviews (those of 4 and 5 stars) received by each item 
	as indicator of its total number of adopters, that potentially might become 
	spreaders of the product, in a situation alike that described by the present model. 
	We analyzed all music albums from Google Play music~\cite{googlemusic}, whose prices are similar (U\$ $9 \pm 3$). 
	We also analyzed male sneakers from Netshoes~\cite{netshoes}, a Brazilian e-commerce for sport goods. 
	Since in this case the prices are more disperse, we split  the data into two subsets: 
	items with prices below and above the median (about R\$ $200$). 
	For each set of data we computed the histogram of the number of adoptions (i.e., number of reviews 
	attributing 4 and 5 stars), as shown in  Fig.~\ref{fig:pdfs-real}.
Comparison of  Figs.~\ref{fig:pdfs_k} and \ref{fig:pdfs-real} 
put into evidence a remarkable qualitative similarity between real and simulated distributions 
producing log-normal shapes. 
The shape for small values of $n_s$ is also similar.
Once  service users have access to the reviews of  any other  user, 
the underlying network  is expected to be similar to  a random graph with relatively high connectivity. 
Since purely random (mean-field) interactions would lead to consensus, which is not observed in 
empirical data, one concludes that the underlying network must have some structure. 
The absence of a fat tail, related to the presence of hubs, as in BA networks, indicates  that the empirical cases 
are best modeled by  ER  networks, at least qualitatively. 
In fact, in a ``rating network'', reviews are equivalent and none of them is expected to act as a hub, 
then, it is reasonable that  ER networks yield more realistic results in this case.

\section{Final remarks}

We introduced a model based on a plurality rule, that mimics decision making  governed by the influence of 
the relative majority of the neighbors. The model  also incorporates the possibility of undecided agents.  
It applies not only to situations involving consume of products or services 
but also to other environments where there is a variety of options,  
as far as the options are homogeneous with similar attractiveness (similar quality, cost, etc.) 
and people have no preferred choice a priori.

Different final steady states emerge from the dynamics, depending on the number of available options 
and on the degree distribution of the network of contacts:  consensus,   wide distribution of opinions or also 
situations where indecision dominates for sufficiently large number of options.  
In fact, decision making  governed by the plurality rule may yield  ties (conflict and frustration), 
contributing to overchoice stagnation.  
This effect appears to be mitigated in BA networks. 
The model envisages  that stagnation may result not only from overchoice but also from the excess of links (see Fig.~\ref{fig:connectivity}.a).

For both types of small-world networks, consensus is  almost certain for sufficiently low number of options 
and sufficiently large connectivity. 
For ER networks, consensus can occur even if there are many options available. If neighbors are random, consensus is the rule. 
But consensus is unlikely for large number of options   and low network connectivity, specially if the 
network is homogeneous. In the square lattice with periodic boundary conditions consensus is not reached, but opinions tend 
to be equipartitioned. 
In small-world networks, there are nontrivial  non-consensus regimes, and a broad distribution of opinions 
can emerge, with a shape similar to that of real ones  
when the assumptions of homogeneity hold, like in the 
examples of Netshoes and Google music albums.

The model indicates that consensus can suddenly emerge simply by  introducing a few connections 
or  eliminating a few items.  
Furthermore, it  also predicts  that an item can become  very popular (with relatively large $f_w$), 
even if the initial attractiveness of all the items is uniform. 
 These observations furnish another possible explanation of why 
there is so much amateur  content viralizing in the Internet, or 
 why a  service, good or cultural product can become a bestseller without having any 
apparent differentiated attractiveness.  

\section*{Acknowledgements}

We acknowledge financial support from Brazilian Agencies CNPq, CAPES and Faperj.

\end{document}